\documentclass{article} 
\usepackage{cite}

\newcommand{\bear}{\begin{eqnarray}}
\newcommand{\eear}{\end{eqnarray}}
\newcommand{\be}{\begin{equation}}
\newcommand{\ee}{\end{equation}}
\newcommand{\beqn}{\begin{eqnarray}}
\newcommand{\eeqn}{\end{eqnarray}}
\newcommand{\beqnn}{\begin{eqnarray*}}
\newcommand{\eeqnn}{\end{eqnarray*}}

\begin{document}

\begin{center} { \bf
UNCERTAINTY RELATIONS FOR TWO OBSERVABLES COUPLED WITH THE THIRD ONE
 } \end{center}

\begin{center} {\bf
V. V. Dodonov 
}\end{center}


\begin{center}
{\it
Institute of Physics and International Center for Physics, University of Brasilia,
P.O. Box 04455, Brasilia 70919-970, Federal District, Brazil}

E-mail:~~~vdodonov@fis.unb.br\\
\end{center}

\begin{abstract}
A new lower boundary for the product of variances of two observables
is obtained in the case, when these observables are entangled with the third one.
This boundary can be higher than the Robertson--Schr\"odinger one.
The special case of the two-dimensional pure Gaussian state is considered as an example.

\end{abstract}


\section{Introduction}	

90 years ago, Heisenberg \cite{Heis} introduced (although in an approximate form) the famous 
``uncertainty relation'' (UR)
\be
\Delta x \Delta p \ge \hbar/2,
\label{dxp}
\ee
that was soon proven 
rigorously in the frameworks of the wave function description of
quantum systems by Kennard \cite{Kennard} and Weyl \cite{Weyl28}. A few years later,
Robertson  \cite{Robertson30} 
 and Schr\"odinger \cite{Schrod30} proved a more general inequality
\be
\sigma_A \sigma_B \ge \frac14\left|
\langle [\hat{A},\hat{B}]\rangle\right|^2
\label{unc3}
\ee
 for arbitrary Hermitian operators $\hat{A}$ and $\hat{B}$.
According to them (and following Heisenberg's idea \cite{Heis}), the "uncertainty" of a quantity A was
defined as the square root of its variance (or mean squared deviation):
\be
\Delta A \equiv \sqrt{\sigma_A}, \qquad
\sigma_A \equiv \langle \hat{A}^2\rangle - \langle \hat{A}\rangle^2.
\label{def-sigA}
\ee

It is known that  inequality (\ref{dxp}) is saturated (becomes the equality) for the wave functions
describing shifted ground states of the harmonic oscillator \cite{Schrodpac}, called nowadays after Glauber
\cite{Glauber} as ``coherent states''.
But what can we say about those quantum states, which possess the left-hand sides of inequalities
(\ref{dxp}) or (\ref{unc3}) bigger than the right-hand ones? One of possible answers is that, 
probably, one should add some extra terms to
the right-hand sides of (\ref{dxp}) or (\ref{unc3}) in such cases, 
taking into account some additional parameters or specific properties
of concrete quantum systems under consideration.
The first step in this direction was made by Robertson  and Schr\"odinger
in the same papers \cite{Robertson30,Schrod30}. 
Namely, they obtained a more precise version of (\ref{unc3}),
taking into account the average value of the anticommutator $\{\hat{A} , \hat{B} \}$:
\be
\sigma_A \sigma_B \ge \sigma_{AB}^2 + \frac14\left|
\left\langle [\hat{A},\hat{B}]\right\rangle\right|^2 \equiv \left|\left\langle (\delta\hat{A})(\delta\hat{B})\right\rangle\right|^2
\equiv G_{AB}^2,
\label{unc4}
\ee
where
\be
\sigma_{AB} \equiv \frac12 \langle \hat{A}\hat{B} + \hat{B}\hat{A}\rangle
- \langle \hat{A}\rangle\langle\hat{B}\rangle
\equiv \frac12 \left\langle \Big\{\delta\hat{A} , \delta\hat{B} \Big\}\right\rangle, 
\qquad \delta\hat{A} \equiv \hat{A} -\langle \hat{A}\rangle.
\label{defsigAB}
\ee
The special case of (\ref{unc4}) is the following generalization of (\ref{dxp}) for the coordinate and momentum operators:
\be
\sigma_p \sigma_x -\sigma_{xp}^2 \ge \hbar^2/4.
\label{unc5}
\ee
The equality takes place for all Gaussian wave functions, as was discovered for the first time by Kennard
\cite{Kennard}.

Inequality (\ref{unc5}) can be rewritten in the form \cite{DKM80}
\be
\sigma_p \sigma_x \ge \frac{\hbar^2}{4\left(1-r^2\right)}, \qquad
r= \frac{\sigma_{xp}}{\sqrt{\sigma_p \sigma_x}},
\label{unc34}
\ee
which emphasizes the role of the ``correlation coefficient'' $r$ as an additional parameter, responsible for the 
increase of product $\sigma_p \sigma_x$. 
One could treat the relation (\ref{unc34}) as
though an ``effective Planck constant'' $\hbar\left(1-r^2\right)^{-1/2}$ is
occurring instead of the usual constant $\hbar$. 
Such an interpretation of inequalities (\ref{unc5}) and (\ref{unc34}) was discussed, e.g., in Refs.
\cite{DKM-200,Vysot13}.
The explicit form of ``correlated coherent states'', saturating inequality (\ref{unc5}), is as follows \cite{DKM80},
\be
\psi(x) =\left(2\pi\sigma_x\right)^{-1/4} \exp\left[-\,\frac{x^2}{4\sigma_x}\left(1-\,\frac{ir}{\sqrt{1-r^2}}\right)
+\frac{\alpha x}{\sqrt{\sigma_x}} -\frac12\left(\alpha^2 +|\alpha|^2\right)\right].
\label{psicorr}
\ee

Inequalities (\ref{unc4}) and (\ref{unc34}) explain the increase of the uncertainty product
$\sigma_A \sigma_B $ due to the existence of some ``intrinsic'' restrictions in the quantum system
under investigation (the nonzero correlation coefficient). 
However, this product can increase also due to some ``extrinsic'' constraints, if the system interacts
with other systems (``environment'').
For example, in the case of an equilibrium state of a harmonic
oscillator with frequency $\omega$ at temperature $T$,
the uncertainty product equals 
$\sigma_{pp}\sigma_{xx}=\left[\frac{\hbar}{2}
\coth\left(\frac{\hbar\omega}{2k_B T}\right)\right]^2 $
($k_B$ is the Boltzmann constant).
In the high-temperature case $k_B T \gg \hbar\omega$ the right-hand side
of this equality is so large, that inequality (\ref{dxp}) 
becomes practically useless.

The equilibrium state of a harmonic oscillator is a {\em mixed\/} quantum states,
described by means of the statistical operator (density matrix) $\hat\rho$.
The degree of mixing is frequently characterized by the difference $1-\mu$, where 
$\mu\equiv \mbox{Tr}(\hat\rho^2)$ is the ``quantum purity''.
It is known that for any quantum state described by means of a
{\em Gaussian\/} density matrix or the Wigner function (in particular, for the equilibrium state), 
the following equality holds
for systems with one degree of freedom (see, e.g., \cite{167}):
\be
 \sqrt{\sigma_{pp}\sigma_{xx} - \sigma_{xp}^2}
=\frac{\hbar}{2\mu}.
\label{SR-mu}
\ee
The generalized ``purity bounded uncertainty relation'' for mixed quantum states can be 
written in the form
\begin{equation}
\sqrt {\sigma_{pp}\sigma_{xx} - \sigma_{xp}^2}\ge\frac {\hbar}2\Phi(\mu),
\label{79}
\end{equation}
where $\Phi (\mu )$ is a monotonous function of $\mu$,
satisfying the relations $\Phi (1)=1\le \Phi(\mu)\le \mu^{-1}$ for $0<\mu\le 1$.
Its explicit form 
turned out rather complicated,
but it can be described with a good accuracy by a simple approximate formula  \cite{183,S98}
\be
\tilde{\Phi }(\mu )=\frac {4+\sqrt {16+9\mu^2}}{9\mu}.
\label{Phitil}
\ee
In particular, the following asymptotical formula holds for $\mu\ll 1$ 
(its leading term was obtained for the first time by Bastiaans \cite{Bast1}):
\begin{equation}
\tilde{\Phi }(\mu )=\frac 8{9\mu}\left(1+\frac 9{64}
\mu^2+\ldots\right),
\label{97}
\end{equation}
so that $|{\Phi} (\mu )-8/(9\mu)|<0,01$ for $\mu\le 0,25$.
Both formulas, (\ref{Phitil}) and (\ref{97}), show that Gaussian states do not minimize the
precise uncertainty relation for mixed states. The minimum value is achieved for some 
diagonal mixtures of finite numbers of the Fock states of the harmonic oscillator.

Inequality (\ref{79}) can be considered as some kind of ``coarse-grained'' relations, since it hides
all details of the interaction (entanglement) between the system under study and the 
``environment''. Our goal is to derive a new inequality, where some of these
details appear explicitly. 

\section{The new inequality and its illustration}
\label{sec-2var3}

The main idea is to start from some inequality related to three observables and find its consequences
with respect to admissible values of the product of two selected variances.
A general scheme of obtaining the uncertainty relations for several observables in terms of covariances was
given by Robertson in 1934 \cite{Robertson34}. Let us remind it.
Consider $N$ arbitrary 
operators $\hat{z}_1$, $\hat{z}_2$, \ldots, $\hat{z}_N$, and construct the operator
$\hat{f} = \sum_{j=1}^N \alpha_j (\hat{z}_j -\langle \hat{z}_j\rangle)$,
where $\alpha_j$ are arbitrary complex numbers.
The inequalities, which can be interpreted as generalized uncertainty relations, 
are the consequences of the fundamental 
inequality $\langle\hat{f}^{\dagger}\hat{f}\rangle\ge 0 $, that
must be satisfied for any pure or mixed quantum
state (the symbol $\hat{f}^{\dagger}$ means the Hermitian conjugated operator).
In the explicit form, this inequality is the condition of
positive semi-definiteness of the
quadratic form $\alpha^*_j F_{jm}\alpha_m $,
whose coefficients
$F_{jm} = \left\langle\Big(\hat{z}_j^{\dagger}- \langle\hat{z}_j\rangle^*\Big)
\Big(\hat{z}_m- \langle\hat{z}_m\rangle\Big)\right\rangle$
form the Hermitian matrix $F =\Vert F_{jm}\Vert$.
One has only to use the known conditions of
the positive semi-definiteness of Hermitian quadratic forms 
to write down the explicit inequalities for the elements of 
matrix $F$. All such inequalities can be considered as
generalizations of inequality (\ref{unc3}) to the case of more than two operators. 
Many of them can be found in the review \cite{183}.

If all operators $\hat{z}_j$ are Hermitian, then it is convenient to
split matrix $F$ as $F = X + iY$, where $X$ and $Y$ are
real symmetric and antisymmetric matrices, respectively,
consisting of the elements
\be
X_{mn}= \frac12 \left\langle\left\{\left(\hat{z}_m-
\langle\hat{z}_m\rangle\right)\,,\,
\left(\hat{z}_n- \langle\hat{z}_n\rangle\right)\right\}\right\rangle,
\qquad
Y_{mn}= \frac1{2i} \left\langle\left[\hat{z}_m\,,\,\hat{z}_n\right]
\right\rangle.
\label{defXY}
\ee
The symbols $\{,\}$ and $[\,,\,]$ mean, as usual, the anticommutator and the
commutator. 
The fundamental inequality ensuring the positive semi-definiteness of matrix $F$ is
$ \det F =\det \Vert X + iY \Vert \ge 0$.
It suits quite well for our purposes, since it contains all elements of matrices $X$ and $Y$.
Its explicit form in the special case of $N=3$ reads
\beqn
X_{11}X_{22}X_{33} &\ge &  X_{11}\left(X_{23}^2 + Y_{23}^2\right)
+ X_{22}\left(X_{13}^2 + Y_{13}^2\right) + X_{33}\left(X_{12}^2 + Y_{12}^2\right)
\nonumber \\ &&
+ 2\left( X_{12}Y_{23}Y_{31} + X_{23}Y_{31}Y_{12} + X_{31}Y_{12}Y_{23} -X_{12}X_{23}X_{31}\right) ,
\label{unc17a} 
\eeqn
Formula (\ref{unc17a}) was obtained by Synge \cite{Synge71}
 (without any reference to Robertson's paper).
Recently, it was re-derived  in Ref.~\cite{Qin16}.

Inequality (\ref{unc17a}) has the form $X_{11}X_{22}X_{33} \ge  a X_{11} + b X_{22} + c $, where
coefficients $a$, $b$ and $c$ do not contain variances $X_{11}$ and $X_{22}$. 
Moreover $a$ and $b$ are non-negative. 
Due to the standard arithmetic-geometric inequality, 
we have $a X_{11}  + b X_{22}  \ge 2\sqrt{a b X_{11} X_{22} }$.
This means that $X_{33}\xi^2 - 2\sqrt{ab}\,\xi -c \ge 0$, where $\xi = \sqrt{X_{11} X_{22}} \ge 0$.
Consequently, $\xi$ must be greater than the biggest root of the quadratic polynomial in the
left-hand side of this inequality:
$X_{33}\xi \ge \sqrt{ab} +\sqrt{ab +cX_{33}}$. Thus we arrive at the inequality
\be
\Delta z_1 \Delta z_2 \ge \sqrt{G_{12}^2 +\Omega^2 +2\Gamma} + \Omega,
\label{2+1}
\ee
where
\be
G_{jk}^2 = X_{jk}^2 +Y_{jk}^2, \qquad
\Omega =\left|G_{13}G_{23}\right|/X_{33}, 
\label{def-Om}
\ee
\be
\Gamma = \left[X_{12}\left(Y_{23}Y_{31} -X_{23}X_{31}\right) 
+  Y_{12}\left(X_{23}Y_{31} +Y_{23}X_{31}\right)\right]/X_{33} .
\label{def-Gam}
\ee
If the observables $z_1$ and $z_2$ are totally independent from $z_3$, then 
$Y_{13}=Y_{23}=X_{13}=X_{23}=0$, and (\ref{2+1}) is reduced to
the Schr\"odinger--Robertson inequality (\ref{unc4}).

If $\left[\hat{z}_1, \hat{z}_3\right] =\left[\hat{z}_2, \hat{z}_3\right]=0$ (for example,  $z_1=x$, $z_2=p_x$ and
$z_3=y$),  then
\be
\Delta z_1 \Delta z_2 \ge \sqrt{Y_{12}^2 +\left(X_{12} - X_{13}X_{23}/X_{33}\right)^2 } 
+\left|X_{13}X_{23}\right|/X_{33}.
\label{2+1comm}
\ee
The right-hand side of this inequality is bigger than the Robertson bound $|Y_{12}|$, if there exist
correlations in the pairs $(z_1,z_3)$ and $(z_2,z_3)$, characterized by nonzero values of the
covariances $X_{13}$ and $X_{23}$.


To illustrate inequality (\ref{2+1comm}), let us consider a special
 two-variable pure Gaussian state, described by the wave function
\be
\psi(x,y)= {\cal N}\exp\left( -\frac{a}{2} x^2 - b xy - \frac{c}{2} y^2\right),
\label{psib}
\ee
where ${\cal N}$ is the normalization factor. 
To simplify the following formulas, let us assume that
coefficients $a$ and $c$ are real (and positive), while $b$ may be an arbitrary complex number, 
satisfying the restriction $D \equiv ac - [\mbox{Re}(b)]^2 >0$.
It is easy to calculate all necessary variances and covariances:
\[
X_{11} \equiv \langle x^2\rangle = \frac{c}{2D}, 
\qquad X_{22} \equiv \langle p_x^2\rangle = \frac{a\hbar^2}{2D}
\left(ac - [\mbox{Re}(b)]^2 + [\mbox{Im}(b)]^2\right),
\]
\[
X_{12} \equiv \frac12\langle \hat{x}\hat{p}_x + \hat{p}_x \hat{x}\rangle =
\frac{\hbar}{2D} \mbox{Re}(b) \mbox{Im}(b), \qquad
Y_{12} = \frac{\hbar}{2},
\]
\[
X_{33} \equiv \langle y^2\rangle = \frac{a}{2D}, \qquad
X_{13} \equiv \langle xy\rangle = -\,\frac{\mbox{Re}(b)}{2D}, \quad
X_{23} \equiv \langle p_x y\rangle = -\,\frac{a\hbar}{2D}\mbox{Im}(b). 
\]
In this case we have $X_{12} = X_{13}X_{23}/X_{33}$, so that (\ref{2+1comm}) can be written
as (here $p \equiv p_x$) $\Delta x \Delta p \ge \hbar/2 +|\sigma_{xp}|$.
The right-hand side of this inequality is certainly bigger than the 
Robertson--Schr\"odinger lower boundary $\left[(\hbar/2)^2 + \sigma_{xp}^2\right]^{1/2}$. 
This happens, because the quantum state describing the  $x$-subsystem is {\em mixed}.
Its density matrix $\rho(x,x^{\prime})= \int \psi(x,y)\psi^*(x^{\prime},y)dy$
has the purity 
$\mu =\left\{\left(ac - [\mbox{Re}(b)]^2\right)/\left(ac + [\mbox{Im}(b)]^2\right)\right\}^{1/2}$.
The equality in (\ref{2+1comm}) is achieved for the states (\ref{psib}) with 
$|\mbox{Re}(b)| = |\mbox{Im}(b)|$.
If $b \neq 0$, then the function (\ref{psib})
cannot be written as a product of some function of $x$ by some function of $y$. In other words,
this function describes an {\em entangled\/} state with respect to variables $x$ and $y$.
Due to this entanglement, the uncertainty product $\Delta x \Delta p$ turns out bigger than the
Robertson--Schr\"odinger boundary.

\label{sec-ex}

\section{Conclusion}

The main results of this paper are inequalities (\ref{2+1}) and (\ref{2+1comm}), which show how the
entanglement of the system under study with other degrees of freedom results in the increase of
the minimal value of
the uncertainty product with respect to the selected system observables. We have found also an
example of quantum states which saturate the new inequality (\ref{2+1comm}). The weakness of
inequality (\ref{2+1comm}) is that it reduces to the Robertson--Schr\"odinger lower boundary (\ref{unc4}),
if one of covariances, $X_{13}$ or $X_{23}$, equals zero, although the uncertanty product
can exceed the boundary (\ref{unc4}) in such cases. Probably, more general and more strict
inequalities can be found, if one applies the scheme of section \ref{sec-2var3} 
to systems of more than three observables. 
This subject is under investigation now.

\section*{Acknowledgments}

A partial support of the Brazilian funding agency CNPq is acknowledged.

\end{document}